\documentclass[12pt]{iopart}
\usepackage{iopams}
\usepackage{epsfig}
\begin{document}

\title[Vortices determine the dynamics of biodiversity in cyclical interactions]{Vortices determine the dynamics of biodiversity in cyclical interactions with protection spillovers}

\author{Attila Szolnoki$^1$ and Matja{\v z} Perc$^{2,3,4}$}
\address{$^1$Institute of Technical Physics and Materials Science, Centre for Energy Research, Hungarian Academy of Sciences, P.O. Box 49, H-1525 Budapest, Hungary\\
$^2$Faculty of Natural Sciences and Mathematics, University of Maribor, Koro{\v s}ka cesta 160, SI-2000 Maribor, Slovenia\\
$^3$Department of Physics, Faculty of Sciences, King Abdulaziz University, Jeddah, Saudi Arabia\\
$^4$CAMTP -- Center for Applied Mathematics and Theoretical Physics, University of Maribor, Krekova 2, SI-2000 Maribor, Slovenia}
\ead{szolnoki@mfa.kfki.hu, matjaz.perc@uni-mb.si}

\begin{abstract}
If rock beats scissors and scissors beat paper, one might assume that rock beats paper too. But this is not the case for intransitive relationships that make up the famous rock-paper-scissors game. However, the sole presence of paper might prevent rock from beating scissors, simply because paper beats rock. This is the blueprint for the rock-paper-scissors game with protection spillovers, which has recently been introduced as a new paradigm for biodiversity in well-mixed microbial populations. Here we study the game in structured populations, demonstrating that protection spillovers give rise to spatial patterns that are impossible to observe in the classical rock-paper-scissors game. We show that the spatiotemporal dynamics of the system is determined by the density of stable vortices, which may ultimately transform to frozen states, to propagating waves, or to target waves with reversed propagation direction, depending further on the degree and type of randomness in the interactions among the species. If vortices are rare, the fixation to waves and complex oscillatory solutions is likelier. Moreover, annealed randomness in interactions favors the emergence of target waves, while quenched randomness favors collective synchronization. Our results demonstrate that protection spillovers may fundamentally change the dynamics of cyclic dominance in structured populations, and they outline the possibility of programming pattern formation in microbial populations.
\end{abstract}

\pacs{87.23.Kg, 89.75.Fb}

\maketitle

\section{Introduction}
Cyclical interactions are at the heart of marine benthic populations \cite{jackson_pnas75}, plant systems \cite{taylor_am90, silvertown_je92, lankau_s07, cameron_jecol09}, and microbial populations \cite{durrett_jtb97, kerr_n02, kirkup_n04, kerr_n06, neumann_gf_bs10, nahum_pnas11}. Cyclic dominance also plays an important role in the overgrowth of marine sessile organisms \cite{burrows_mep98}, the mating strategy of side-blotched lizards \cite{sinervo_n96}, the genetic regulation in the repressilator \cite{elowitz_n00}, and in explaining the oscillations of the population size of lemmings \cite{gilg_s03} and the Pacific salmon \cite{guill_jtb11}. More generally, evolutionary games entailing cyclic dominance play a prominent role in explaining biodiversity \cite{frachebourg_prl96, durrett_tpb98, frean_prsb01, czaran_pnas02, reichenbach_n07, lutz_jtb13, mobilia_jtb10, laird_oikos14, kelsic_n15, rulquin_pre14, bergstrom_n15}, and they are also able to provide insights into Darwinian selection \cite{maynard_n73}, structural complexity \cite{watt_je47}, and prebiotic evolution \cite{rasmussen_s04}, as well as into the effectiveness of positive and negative reciprocity \cite{szolnoki_prx13}, volunteering \cite{hauert_s02, semmann_n03}, rewarding \cite{hauert_jtb10, szolnoki_epl10}, and punishment \cite{hauert_s02, sigmund_n10, helbing_ploscb10, amor_pre11, szolnoki_pre11}, to name but a few representative examples.

In agreement with the impressive implications fundamental research on cyclical interactions has, it is little surprising that the classical rock-paper-scissors game -- the workhorse for research on cyclic dominance -- has been studied so extensively, not least by methods of statistical physics \cite{szabo_pre99, mobilia_pre06, reichenbach_prl07, peltomaki_pre08, berr_prl09, he_q_pre10, wang_wx_pre10b, mowlaei_jpa14, wang_wx_pre11, mathiesen_prl11, avelino_pre12, roman_jsm12, avelino_pre12b, juul_pre12, roman_pre13, rulands_pre13, vukov_pre13, groselj_pre15, intoy_pre15}, which are indispensable for a comprehensive treatment of the game and its extensions in structured populations. Although the rules of the game can be written down in a short sentence, the complexity of spatial patterns that emerge spontaneously as a consequence of the simple microscopic rules is unparalleled. Labyrinthine clustering \cite{juul_pre13} and interfaces with internal structure \cite{avelino_pre14} are just two of the most intriguing recent examples attesting to this fact.

Despite the overwhelming attention the rock-paper-scissors game has received \cite{szolnoki_jrsif14}, there is one important aspect of the game that has till recently been overlooked. In particular, while the direction of invasion between the rock, the scissors and the paper is intransitive, and thus conductive to species coexistence \cite{laird_an06, laird_jtb15}, this may not be the case for the protection one species offers to the other. If one considers the game not from the standpoint of dominance in that rock is wrapped by paper, paper is cut by scissors, and scissors are broken by a rock, but rather from the standpoint of survival in that scissors protect itself from paper's toxin, paper protects itself from rock's toxin, and the rock protect itself from scissors's toxin, then one quickly comes to realize that such protection may be non-excludable, and that in fact it can spill over to the other strategy. Motivated by this important consideration, Kelsic et al. \cite{kelsic_n15} have recently introduced cyclical interactions with protection spillovers as a new paradigm for biodiversity in well-mixed microbial populations. By using simulations and analytical models, they have show that the opposing actions of antibiotic production and degradation enable coexistence even if the interactions among the cyclically dominating species are random. In a commentary to the original paper, Bergstrom and Kerr \cite{bergstrom_n15} have generalized these results to the classical rock-paper-scissors game in a well-mixed population, revealing a stable, attractive equilibrium containing all three species if the possibility is given that a predator's predator can protect the prey of the former.

However, since interactions in microbial populations are often not random \cite{kerr_n02, czaran_pnas02, nahum_pnas11}, it is important to determine the merits of protection spillovers also in structured populations. We therefore study the rock-paper-scissors game with protection spillovers on the square lattice with annealed and quenched randomness. In a structured population, the sole presence of rock might prevent scissors from beating paper, simply because rock beats scissors. As we will show, this seemingly small modification of the microscopic dynamics has rather spectacular consequences for the collective behavior of the system. Unlike in the classical rock-paper-scissors game, here the spatiotemporal dynamics is determined by the density of stable vortices, which may ultimately transform to frozen states, to propagating waves, or to target waves with reversed propagation direction. Since the initial density of vortices is controllable in experimental setups, like for example in a Petri dishe \cite{kerr_n02} or in bacterial biofilms \cite{drescher_cb14}, our results thus reveal a feasible way of programming pattern formation in microbial populations. Different from the classical rock-paper-scissors game, where we can observe globally synchronized oscillations in the frequency of strategies as we increase the level of randomness (either quenched or annealed), here the consideration of these two different types of randomness can be a decisive factor that determines the evolutionary outcome. As we will show, annealed and quenched randomness have a completely different impact on the emergence of stable spatial patterns, thus demonstrating that protection spillovers may fundamentally change the spatiotemporal dynamics of cyclic dominance in structured populations.

The organization of this paper is as follows. We present the definition of the spatial rock-paper-scissors game with protection spillovers and the details of the Monte Carlo simulation procedure in Section~II. Main results are presented in Section~III. We conclude with the summary of the results and a discussion of their implications in Section~IV.

\section{Rock-paper-scissors with protection spillovers}
We consider a modified version of the classic rock-paper-scissors game, where the three species cyclically dominate each other. For convenience, we refer to the species as $R$, $P$ and $S$, where strategy $R$ invades strategy $S$, strategy $S$ invades strategy $P$, and strategy $P$ invades strategy $R$. However, due to the consideration of protection spillovers, these invasions occur only if none of the direct neighbors of the prey is a predator to the original predator. For example, rock is unable to invade scissors if one of the direct neighbors of the scissors is paper. Similarly, scissors are unable to invade paper if one of the direct neighbors of the paper is rock, and paper is unable to invade rock if one of the direct neighbors of the rock are scissors.

\begin{figure}
\centerline{\epsfig{file=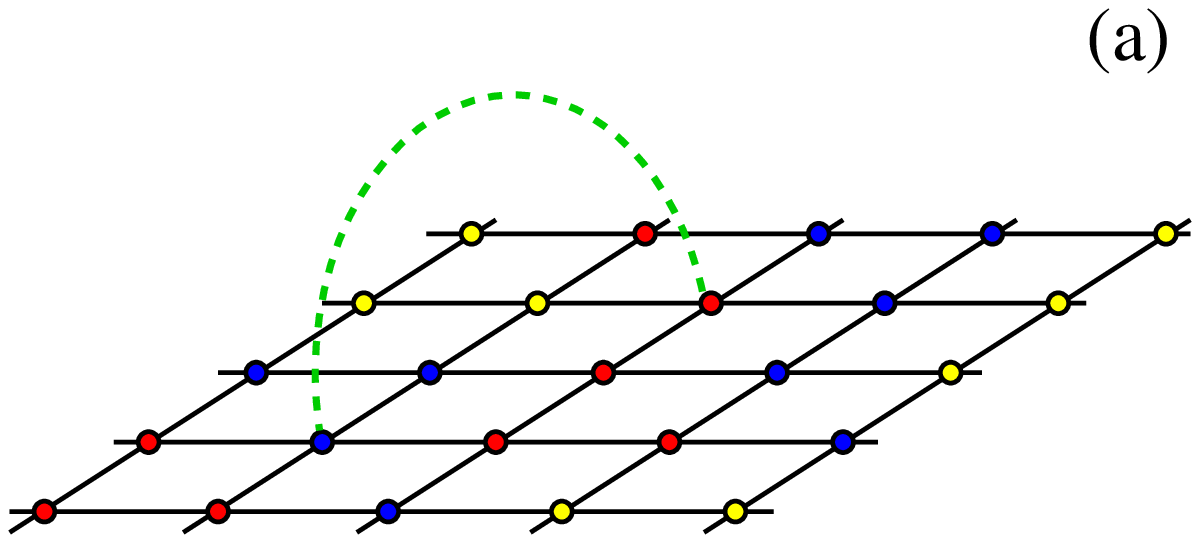,width=8cm}\epsfig{file=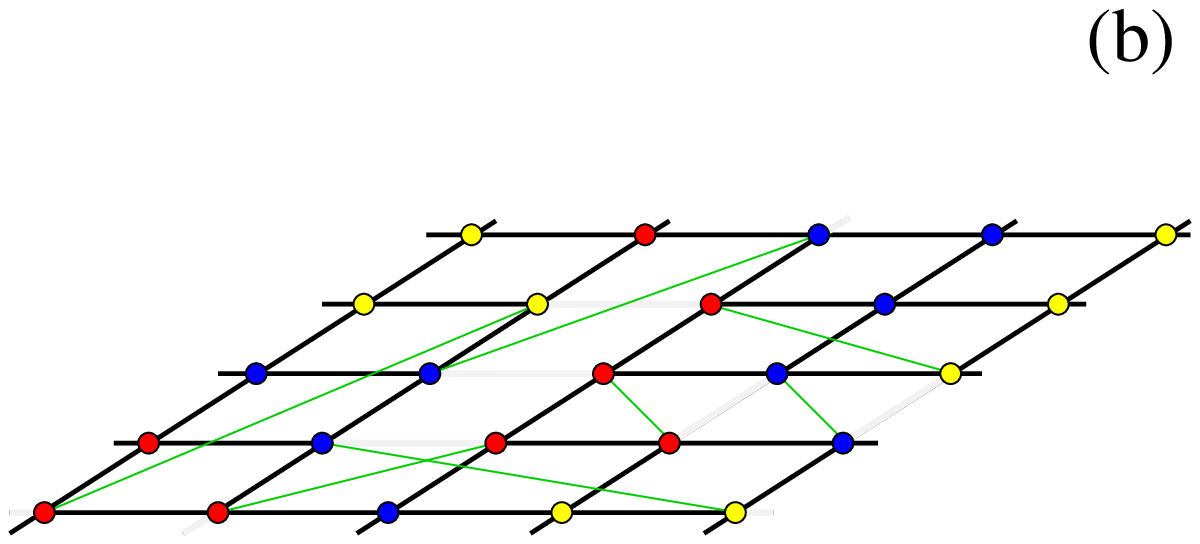,width=8cm}}
\caption{\label{struct} Schematic illustration of annealed (left) and quenched (right) randomness. Annealed randomness preserves the regular interaction structure, but a player may invade a randomly chosen distant player with probability $\vartheta$ (denoted by a dashed green line), while with probability $1-\vartheta$ the invasion remains bounded to nearest neighbors. Quenched randomness requires that a fraction $\theta$ of nearest-neighbor links (denoted by light grey lines) is rewired randomly (denoted by thin green lines), all the while preserving the degree of every node. This interaction structure does not change over time, and so the invasions can occur only via the partly (depending on the value of $\theta$) randomized interaction network.}
\end{figure}

The described rock-paper-scissors game with protection spillovers is studied in structured populations. Each species is thus located on the site $x$ of a square lattice with periodic boundary conditions, where the grid contains $L \times L$ sites. In addition, we also explore the impact of disorder, which has proven to be a decisive factor in the classical rock-paper-scissors game \cite{szabo_jpa04, szolnoki_pre04b}. In particular, the introduction of structural randomness can trigger a global oscillatory state, which is impossible to observe in the absence of long-range interactions. Interestingly, previous research on the classical rock-paper-scissors game \cite{szabo_jpa04, szolnoki_pre04b} has emphasized that the type of disorder by means of which long-range links are introduced has only second-order importance, given that both annealed and quenched randomness have a qualitatively similar impact on pattern formation. However, the introduction of protection spillovers may significantly affect the dynamics of cyclic dominance \cite{kelsic_n15}, which is why we here consider the impact of annealed and quenched randomness separately. As panel (a) of Fig.~\ref{struct} illustrates, annealed randomness is introduced so that at each instance of the game a potential target for an invasion is selected randomly from the whole population with probability $\vartheta$, while with probability $1-\vartheta$ the invasion is restricted to a randomly selected nearest neighbor \cite{szabo_jpa04, szolnoki_pre04b}. For $\vartheta=1$ we thus obtain well-mixed conditions, while for $\vartheta=0$ only short-range invasions along the original square lattice interaction structure are possible. Quenched randomness, illustrated in panel (b) of Fig.~\ref{struct}, is introduced by randomly rewiring a fraction $\theta$ of the links that form the square lattice whilst preserving the degree of each site. We thereby obtain regular small-world networks for small values of $\theta$ and a regular random network in the $\theta \to 1$ limit. Importantly, the rewiring is performed only once before the start of the game, thus introducing quenched (time invariant) randomness in the interactions among the species.

The evolution of species proceeds in agreement with a random sequential update, where during a full Monte Carlo step (MCS) every player receives a chance once on average to invade one randomly selected neighbor (or any member of the population with probability $\vartheta$ in case annealed randomness is considered), as allowed by the rules of the game. The average fraction of rock ($\rho_{R}$), paper ($\rho_{P}$), and scissors ($\rho_{S}$) in the population is monitored during the whole evolutionary process. We have systematically applied different system sizes ranging from $L=40$ to $L=1000$ to reveal the possible size-dependence of the observed solutions. When determining the fixation probability, we have averaged the outcome over $10000$ independent runs. The monitoring time, which exceeded $10^7$ MCS for the largest system size, was always chosen to be at least $100$ times longer than the longest measured fixation time.

\section{Results}
Before presenting the main results, we briefly contemplate on the potential impact of protection spillovers in cyclical interactions. Foremost, it is important to note that the introduction of protection spillovers raises an interesting dilemma that is otherwise absent in the classical version of the rock-paper-scissors game. Namely, the ``spillover protector'' of a species is simultaneously also its prey. Thus, each time an invasion is made, the invading species potentially (although not certainly because there may be other instances of the same prey in the neighborhood) looses the benefit of spillover protection. The predator is thus faced with a difficult choice. Perhaps even more frustratingly, the predator is unable to actually make a choice. The invasion will go forward with certainty if only the prey is not protected by a ``spillover protector''. This dilemma is a good indicator of the fact that a correct intuitive anticipation of the impact of protection spillovers is not at all trivial.

Regarding the potential impact of annealed and quenched randomness, previous research has revealed that both sources of randomness have the same impact on cyclical interactions in that they evoke synchronized oscillations among the competing species \cite{szabo_jpa04, szolnoki_pre04b}. As we will show in the following subsections, this conceptual similarity no longer exists if protection spillovers are introduced.

\subsection{Evolution on the square lattice}

\begin{figure}
\centerline{\epsfig{file=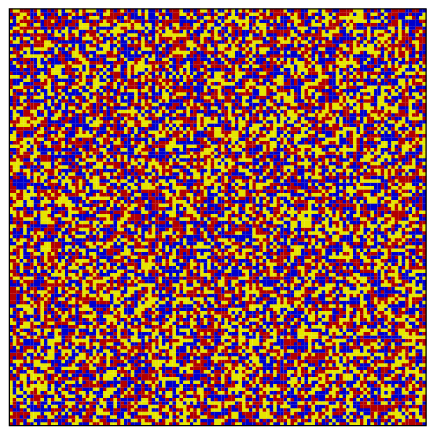,width=4.25cm}\epsfig{file=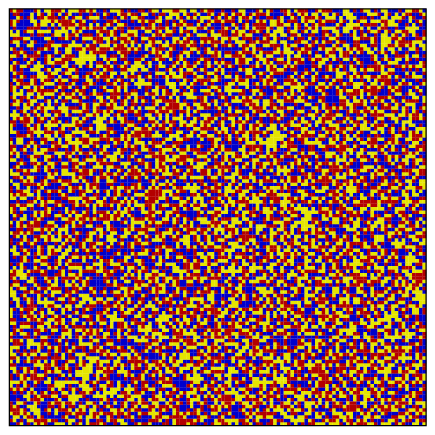,width=4.25cm}}
\centerline{\epsfig{file=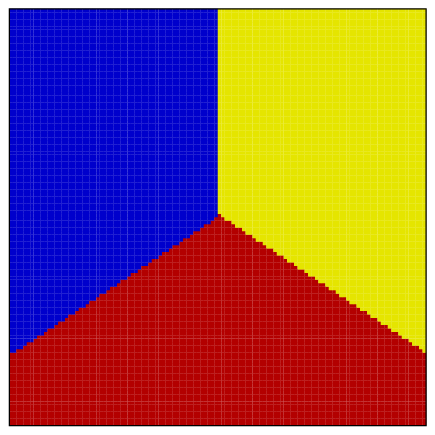,width=4.25cm}\epsfig{file=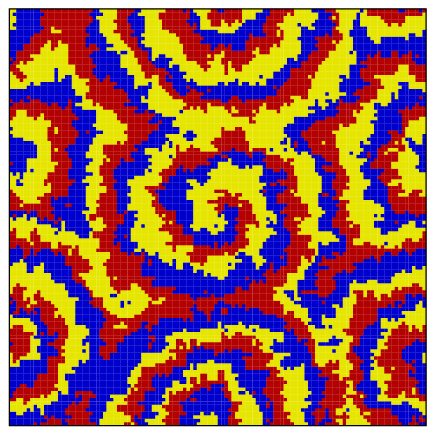,width=4.25cm}}
\caption{\label{two_outcomes} Snapshots of the square lattice, showing a characteristic evolution as obtained from a random initial state (top row) and from a prepared initial state (bottom row). If initially all three species are distributed uniformly at random, as shown in the top left panel, the system quickly evolves to an almost unchanged frozen state (top right panel) where invasions no longer occur. However, if we use special initial conditions where initially there are only two vortex-antivortex pairs (one in the middle and the rest at the edge of the lattice due to periodic boundary conditions), as shown in the bottom left panel, the system evolves into an active stationary state that is characterized by propagating waves (bottom right panel). These propagating waves are much the same as those we can observe in the classic rock-paper-scissors game. Interestingly, in all four depicted snapshots it holds that $\rho_R = \rho_P = \rho_S = 1/3$, yet the stationary realization of this biodiversity is lastly very different. For clarity, we have here used a small square lattice with linear size $L=120$.}
\end{figure}

We begin by presenting the main results obtained on the square lattice in the absence of both annealed and quenched randomness. The first key fact is illustrated in the four snapshots that are depicted in Fig.~\ref{two_outcomes}, which is that the initial state determines significantly the final outcome of the game. More precisely, if the rock-paper-scissors game with protection spillovers is launched with random initial conditions, as shown in the top left panel, then the population terminates rather quickly in a frozen state, as shown in the top right panel. It can be observed at a glance that both the configuration and the density of the three competing species remain practically unchanged during the evolution. Of course, once the frozen state is reached, the invasions seize completely. On the other hand, if the game is initiated with prepared initial conditions, as shown in the bottom left panel of Fig.~\ref{two_outcomes}, then the propagating waves emerge that are qualitatively identical to those that have been observed so often in the classical spatial rock-paper-scissors game \cite{szolnoki_jrsif14} (for the corresponding movie see \cite{movie1}). The crucial property that characterizes the special initial condition shown in the bottom left panel is that it contains two vortex-antivortex pairs, one vortex in the middle of the lattice, and the rest of the vortices located at the edge of the lattice due to periodic boundary conditions. These vortices act as sources for the propagating waves, as can be deduced clearly from the stationary state that is depicted in the bottom right panel of Fig.~\ref{two_outcomes}. We emphasize that $\rho_R = \rho_P = \rho_S = 1/3$ holds for all four depicted snapshots, and that thus all four states are represented by the same point in a simplex. Evidently, this illustrates that the representation of the actual state in a simplex is not always satisfactory for a spatial system. But more importantly, the equality of the density of species in both stationary states indicates that the biodiversity in the rock-paper-scissors game with protection spillovers can be realized in very different ways.

\begin{figure}
\centerline{\epsfig{file=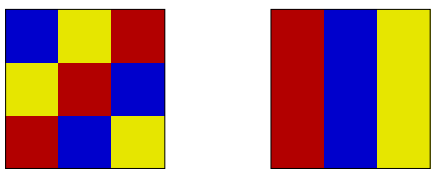,width=6cm}}
\caption{\label{3} Two possible realizations of a frozen state, as obtained on a $L \times L = 3 \times 3$ square lattice with periodic boundary conditions. Due to the introduction of protection spillovers, the depicted configuration of species in both cases mutually prevents successful invasions of other species. These tilted (left) and axial (right) stripes are the fundamental building blocks of frozen states in large populations.}
\end{figure}

To better understand and explain the two significantly different evolutionary outcomes depicted in Fig.~\ref{two_outcomes}, we present in Fig.~\ref{3} two simple $L \times L = 3 \times 3$ configurations in which there is no invasion between the species due to the mutual protection between predator-prey pairs. If we check the depicted configurations carefully, it can be observed that a predator cannot invade a neighboring prey because there is always a predator's predator in the neighborhood of the prey, and who thus protects the prey due to the consideration of protection spillovers. These configurations represent axial or tilted stripes, which are actually the building blocks of frozen states also in much larger populations. Evidently, a ``random'' distribution of species can also yield a frozen state, as it is illustrated in the top row of Fig.~\ref{two_outcomes}, but such an outcome requires many vortices be present in the initial distribution of the species (or a fully random initial distribution, as is depicted in the top left panel of Fig.~\ref{two_outcomes}).

\begin{figure*}
\centerline{\epsfig{file=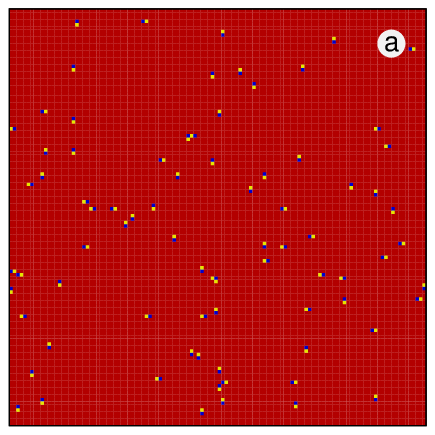,width=3.9cm}\epsfig{file=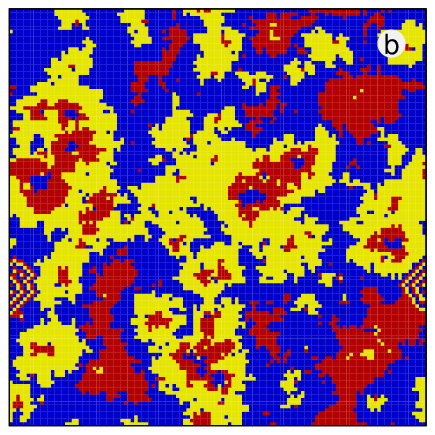,width=3.9cm}\epsfig{file=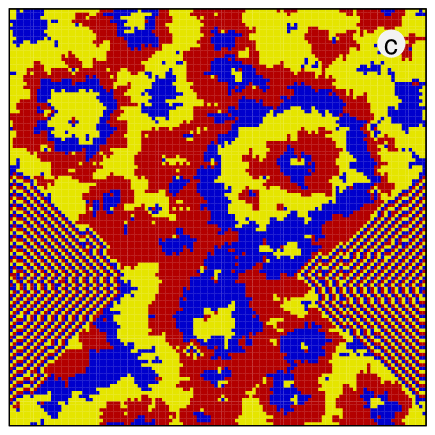,width=3.9cm}\epsfig{file=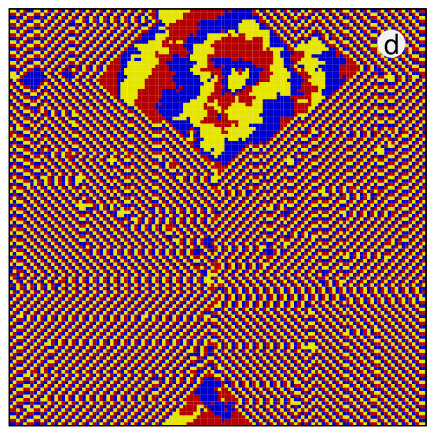,width=3.9cm}}
\caption{\label{terminate_to_frozen} Snapshots of the square lattice, showing a characteristic evolution towards a frozen state from left to right, as obtained from an initial configuration containing several vortices. While each vortex is potentially a source of propagating waves, once these waves emerge they collide, eventually leading to the formation of a relatively small frozen patch (see panel b), which ultimately grows to occupy the entire lattice. For clarity, we have here used a small square lattice with linear size $L=120$, with $80$ vortices inserted at random. Technically, we have inserted $80$ pairs of the two species (yellow and blue) that are different from the ``background'' species (red). Here the frozen state is reached within $600$ MCS.}
\end{figure*}

The crucial role of the initial density of vortices for the final outcome of the game is illustrated in Fig.~\ref{terminate_to_frozen}. Unlike in the bottom left panel of Fig.~\ref{two_outcomes}, where only a few vortices are initially present, we now initiate the game with many more vortices initially present in the population. As panel~(a) of Fig.~\ref{terminate_to_frozen} shows, we initially have a homogeneous state in which $80$ vortices are inserted uniformly at random. These vortices all serve as potential sources of propagating waves, and indeed many do nucleate (for the corresponding movie see \cite{movie2}). The critical point of evolution occurs when a relatively small frozen patch emerges, as illustrated in panel~(b). We emphasize that the pattern in the frozen patch is initially identical to the one that is depicted in the left panel of Fig.~\ref{3}. In fact, such frozen patches typically emerge when propagating wave collide and give rise to sizable patches that are made up of stable patterns shown in Fig.~\ref{3}. During the course of evolution the area that is occupied by such stable patches grows, as shown in panel~(c), while the area where invasions are still possible shrinks and becomes limited to smaller and smaller adjacent domains. Eventually even the last remaining active areas, as shown in panel~(d), become frozen to yield the final frozen state (not shown in Fig.~\ref{3}, but can be seen in the corresponding movie \cite{movie2}). The latter is made up of a mixture of fundamental frozen patterns that are shown in Fig.~\ref{3}

\begin{figure}
\centerline{\epsfig{file=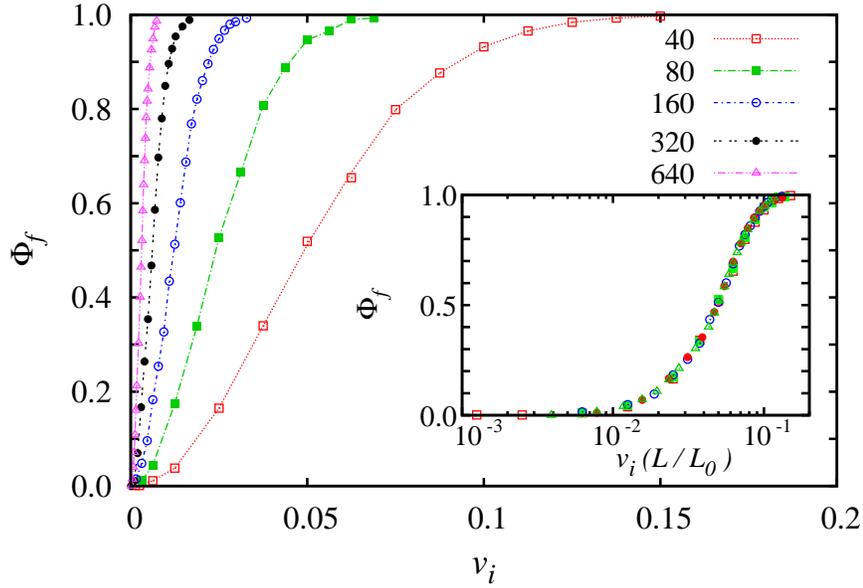,width=12.0cm}}
\caption{\label{frozen_lattice} The probability to reach the frozen state ($\Phi_f$) in dependence on the initial density of vortices ($v_i$), as obtained for different system sizes (see legend). It can be observed that the smaller the system size, the more frequent initially vortices need to be for the propagating waves to eventually terminate in a frozen state. The inset shows the scaled version of the curves depicted in the main panel to the reference system size $L_0=40$. Depicted results are averages over $10^4$ independent runs.}
\end{figure}

Evidently, the emergence of viable frozen domains that are able to grow depends sensitively not just on the number, but also on the initial distribution of the vortices. The final outcome could be different even if we start with the same number of vortices, because their proximity on the lattice could be a decisive factor as well. Accordingly, the emergence of frozen states is a stochastic process. The probability to reach the frozen state ($\Phi_f$) in dependence on the initial density of vortices ($v_i$) is illustrated in Fig.~\ref{frozen_lattice}, while the inset shows the scaled version of curves to the reference system size $L_0=40$. The depicted results indicate that the population becomes more and more sensitive to the initial presence of vortices as we increase its size. More precisely, the initial density of vortices required to reach fixation decreases as we increase the system size. This counterintuitive phenomenon is related to the application of periodic boundary conditions. In particular, the vortices serve as permanent sources of propagating waves, and if the system size is small, then a specific source can easily interact with its virtual ``clone'' due to periodic boundary conditions. The real and virtual sources emit waves synchronously which prevents blocking of akin propagating waves. Accordingly, we need a relatively larger number of vortices to reach fixation here. Naturally, this effect becomes weaker as we enlarge the system size, and thus fewer initial vortices suffice to reach the same fixation probability. On the other hand, we note that a ``non-frozen'' state actually corresponds to a stationary state with endlessly emerging propagating waves, rather than a failure on our side to wait sufficiently long for the system to get trapped in a frozen state. Just to illustrate this fact, a system with linear size $L=80$ size remains in the active state up to $4\times10^6$ MCS. While there is of course a fluctuation about the fixation time having normal distribution, the monitoring trial time was always at least $100$ times longer than the longest observed fixation time.

\subsection{Evolution on the square lattice with annealed randomness}
We first consider the impact of annealed randomness on the dynamics of biodiversity in the studied rock-paper-scissors game with protection spillovers. As described in Section~II, during each instance of the game there is thus a probability $\vartheta$ that a potential target for invasion will be selected from the whole population rather than from the nearest neighbors of a square lattice. In the classical rock-paper-scissors game, the introduction of annealed randomness evokes a synchronized state where the densities of species oscillate in time (note that these oscillations are absent in the absence of annealed randomness). The amplitude of these oscillations increased as we increase $\vartheta$, and at a critical value the system terminates into an absorbing phase where a single species survives \cite{szabo_jpa04, szolnoki_pre04b}.

In the rock-paper-scissors game with protection spillovers the impact of annealed randomness is significantly different. If we start with an initial condition where all three species are distributed uniformly at random, then the population again, as on the square lattice, terminates rather quickly in a frozen state (see top row of Fig.~\ref{two_outcomes}). Here the value of $\vartheta$ plays a negligible role. In exploring other initial conditions, we find that the key factor is again the density of vortices that are initially present in the population. If the vortices are sufficiently common the frozen state is practically unavoidable, and indeed very similar fixation curves can be obtained as we have shown for the original square lattice topology in Fig.~\ref{frozen_lattice}. This behavior can be observed regardless of the strength of annealed randomness $\vartheta$. However, if the vortices are initially rare, the non-frozen state is significantly different from the propagating waves depicted in the bottom right panel of Fig.~\ref{two_outcomes}.

\begin{figure*}
\centerline{\epsfig{file=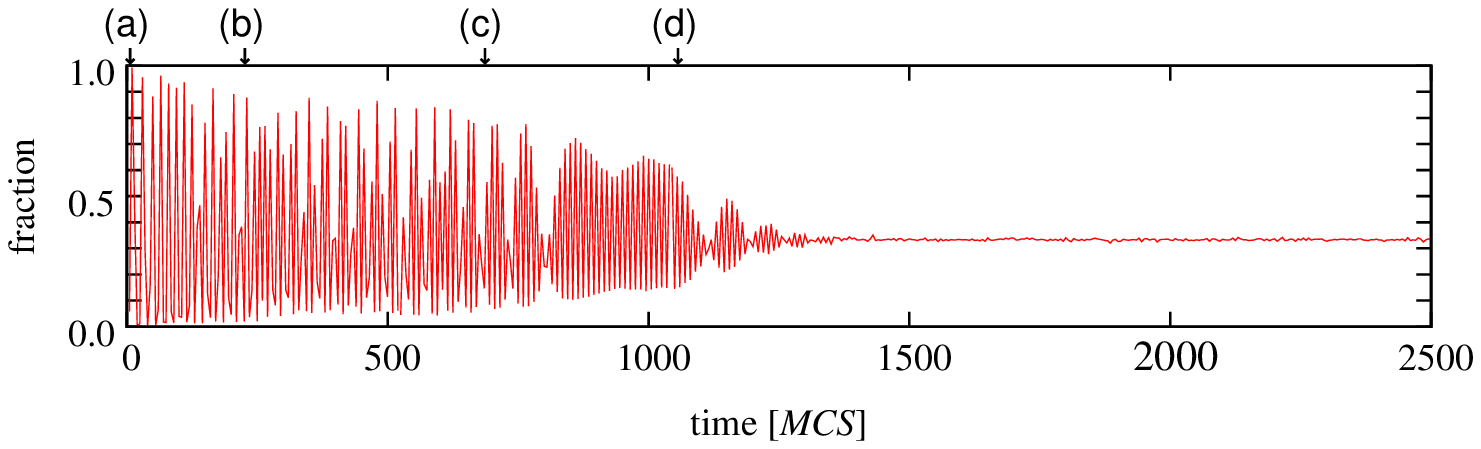,width=14cm}}
\centerline{\epsfig{file=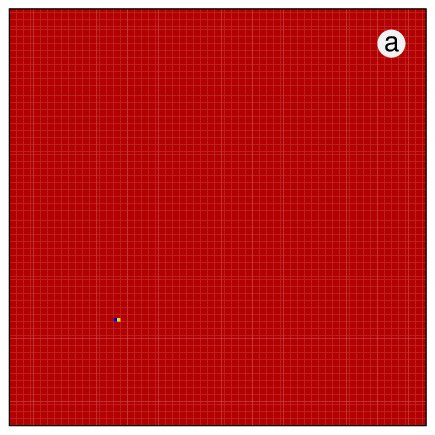,width=3.9cm}\epsfig{file=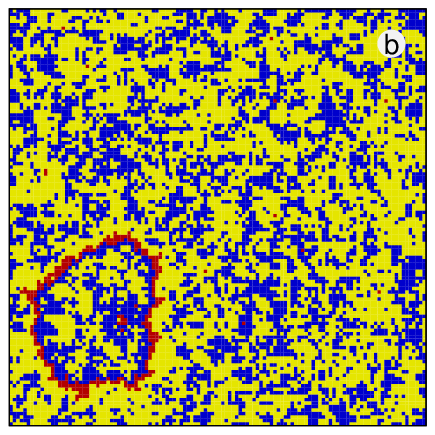,width=3.9cm}\epsfig{file=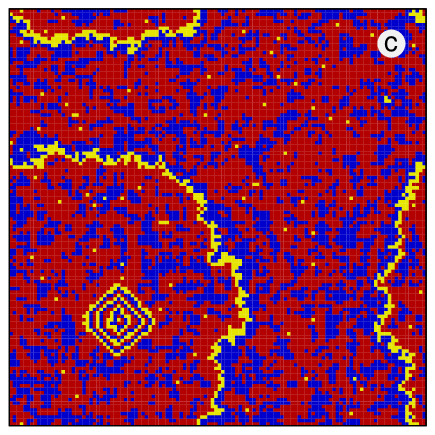,width=3.9cm}\epsfig{file=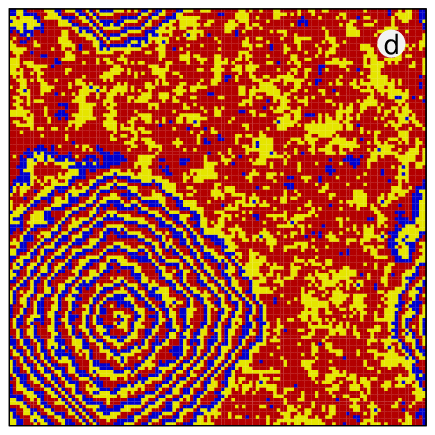,width=3.9cm}}
\caption{\label{single} The time evolution of the density of an arbitrary species (top panel), and the corresponding snapshots of the square lattice (bottom row), showing a characteristic evolution towards target waves with reversed propagation direction. These waves are practically moving backward (towards the center), while the area occupied by these concentric waves expands gradually. The only vortex that is initially located in the middle of the bottom left quadrant of the lattice (see panel a) ultimately acts as the sink for the waves, while the location where the wavefronts collide due to the periodic boundary conditions in the top right quadrant of the square lattice acts like the source of the wavefronts. The stationary state is not frozen as the time course in the top panel might suggest. Instead, wavefronts emerge continuously at the source and travel towards the sink, in the direction that is opposite to the impression one might have if simply looking at the snapshot from left to right. The strength of annealed randomness is $\vartheta=0.22$, and for clarity, we have again used a small square lattice with linear size $L=120$.}
\end{figure*}

To illustrate what happens and to explain the origin of the difference, we monitor the evolution at a representative value of $\vartheta=0.22$ from a prepared initial state, where only a single vortex is initially present in the population (for the corresponding movie see \cite{movie3}). The related Fig.~\ref{single} shows how the fraction of an arbitrary species (top panel) and the spatial distribution of species (bottom row) evolve. In panel~(a) the vortex is initially placed at the center of the bottom left quadrant of the square lattice. Similarly as in the random-free case, this vortex first serves as a source of propagating waves. The expansion of the initial waves can be observed in panels~(b) and (c). The background, however, changes intensively, which is the well-known consequence of synchronization that emerges due to the introduction of annealed randomness. A crucial difference here is that the presence of the vortex does not allow the system to terminate into a homogeneous, absorbing phase. We note that the system would in fact be in an absorbing phase at such a high value of $\vartheta$ in the classical rock-paper-scissors game, as illustrated in Fig.~5 of Ref.~\cite{szabo_jpa04}. In the present case, however, the vortex behaves like a fixed point whose position does not change in time because neither nearest nor distant neighbors are able to invade the species who form the vortex due to protection spillovers that warrant them mutual immunity.

As time goes by, target waves emerge as illustrated in panel~(c) of Fig.~\ref{single}. It is easy to see that the configuration of species within these waves is practically identical to the fundamental building blocks of frozen states that we have shown in Fig.~\ref{3}. Accordingly, this ``stripe-like'' targets are stable against the invasion attempts of far-away other species, which happens frequently at nonzero values of $\vartheta$. Interestingly, and unlike in the absence of annealed randomness, however, these target waves do not represent frozen states. Instead, the depicted target waves propagates \textit{towards} the vortex in the bottom left quadrant of the lattice. When they arrive at the vortex they simply vanish. More precisely, the vortex permanently erodes the nearest locally homogeneous wavefront, which results in a continuous shift of the stripes toward the center of the vortex. The vortex thus acts like a sink for the target waves, and its position never changes over time. We note that the reversed propagation direction of the target waves can be inferred from the snapshots because the order of colors, and hence of the species, is different from the order in the propagating front. This ``dynamically stable'' domain around the vortex will growth until it reaches the frontier of a similar domain (in panel~(e) the domain actually meets itself due to the periodic boundary conditions). In contrast to the random-free case, here the area in the middle of the upper right quadrant of the lattice, where there is a ``void'' between the target wave domains, changes color periodically due to the invasions from distant sites (which is made possible by annealed randomness), and thus acts like the source of the waves. The wavefronts travel outward to join the target wave domains, and then continue towards the center of the vortex which acts as their sink. In even larger populations, it is possible that more than one such source-sink pairs emerge, resulting in a dynamically stable pattern where the frequencies of all three species are equal at all times. A representative evolution for the latter case can be seen in the movie~\cite{movie4}.

\subsection{Evolution on the square lattice with quenched randomness}

\begin{figure}
\centerline{\epsfig{file=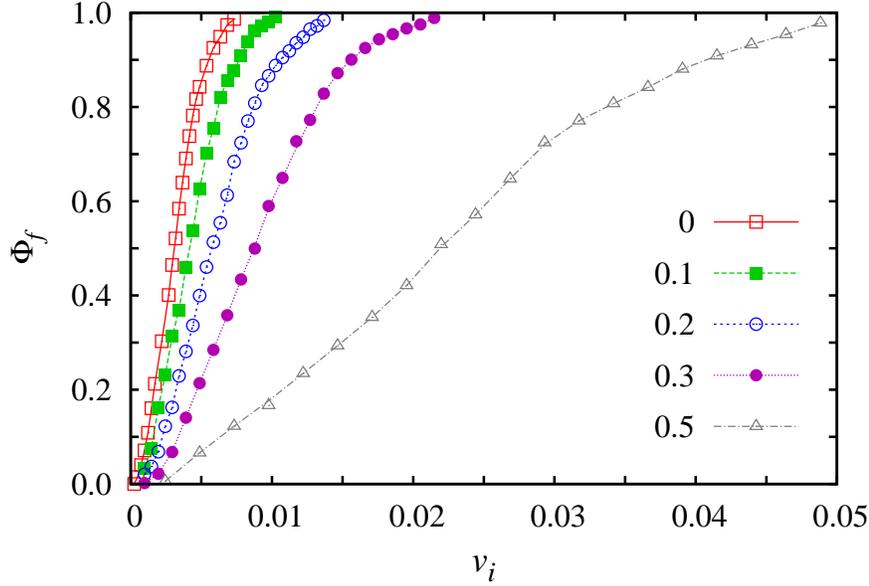,width=12cm}}
\caption{\label{fixation_Q} The probability to reach the frozen state ($\Phi_f$) in dependence on the initial density of vortices ($v_i$), as obtained for different values of $\theta$ that determine the strength of quenched randomness (see legend). It can be observed that the more the square lattice approaches the regular random graph limit ($\theta \to 1$), the more frequent initially vortices need to be for the propagating waves to eventually terminate in a frozen state. In other words, the higher the randomness in the interactions, the more vortices are initially allowed for the system to still avoid a frozen state. Depicted results are averages over $10^4$ independent runs with a linear system size $L=640$.}
\end{figure}

An alternative way of introducing randomness in the interactions among the species is by randomly rewiring a fraction $\theta$ of the links that form the square lattice, whereby we obtain regular small-world networks for small values of $\theta$ and a regular random network in the $\theta \to 1$ limit \cite{szabo_jpa04, szolnoki_pre04b}. Since the interaction topology does not change over time, this setup corresponds to quenched randomness.

As in all previously considered cases, here too a random initial state terminates quickly in a frozen state, and the initial fraction of vortices in the population is a key determinant of the spatiotemporal dynamics that subsequently emerges. The more frequent the vortices, the likelier the population will terminate in a frozen state. Unlike by annealed randomness, however, here the value of $\theta$ plays an important role in determining the conditions that lead to fixation. This fact is illustrated in Fig.~\ref{fixation_Q}, where we show the fixation probability in dependence on the initial density of vortices for different values of $\theta$. The presented results indicate that the less ordered the interaction structure, the less likely is the fixation to a frozen state. Importantly, we emphasize that, even in the $\theta=1$ limit, when the strength of quenched randomness is maximal, completely random initial conditions will still inevitably fixate to a frozen state.

On the other hand, if initially the vortices are only few and the system is hence able to avoid fixation, then the resulting spatiotemporal dynamics is significantly different from the one depicted in Fig.~\ref{single} for annealed randomness. Similarly as in the classical version of the rock-paper-scissors game without protections spillovers \cite{szabo_jpa04}, in this case collective synchronization among the species emerges, yielding oscillations of their densities, or equivalently, a periodic orbit in the ternary diagram. The introduction of protection spillovers thus does not qualitatively modify the impact of quenched randomness on the dynamics of the spatial rock-paper-scissors game, if only the population can avoid fixation to a frozen state.

\begin{figure}
\centerline{\epsfig{file=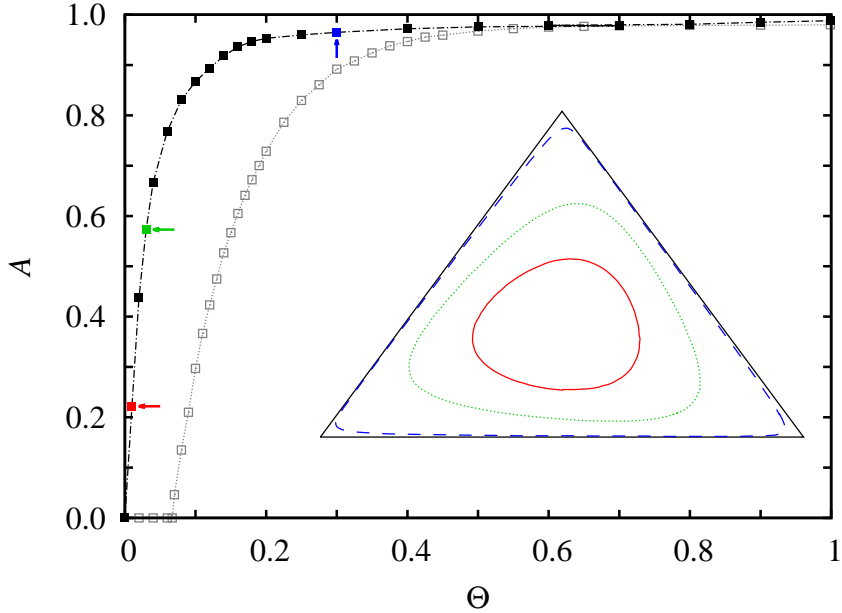,width=12cm}}
\caption{\label{few_vortex} The area of closed orbits ($A$) that correspond to stationary non-frozen states in the ternary diagram in dependence on the strength of quenched randomness ($\theta$), as obtained for the classical rock-paper-scissors game (open squares) and the rock-paper-scissors game with protection spillovers (filled squares). In the latter case, we have used the initial density of vortices $v_i=0.001$ at a linear system size $L=1000$. Although the impact of quenched disorder is in principle similar in both studied versions of rock-paper-scissors game in that it promotes the emergence of collective synchronization in the system, the shift towards larger values of $A$ demonstrates clearly that protection spillovers further enhance this effect. The inset shows the corresponding orbits in the ternary diagram, as obtained at $\theta=0.01$ (solid red line), $\theta=0.03$ (dotted green line), and at $\theta=0.3$ (dashed blue line) in the rock-paper-scissors game with protection spillovers. The areas of these orbits are marked by arrows with the same color in the main panel.}
\end{figure}

Nevertheless, protection spillovers do affect the stationary state in that sufficiently rare vortices facilitate the emergence of collective synchronization. This effect is illustrated in Fig.~\ref{few_vortex}, where we compare the level of synchronization in the rock-paper-scissors game, as obtained without and with protection spillovers. In the latter case the applied initial density of vortices is low ($v_i=0.001$), where fixation to a frozen state is thus very unlikely. The presented results indicates clearly that even a minute fraction of long-range links evokes a synchronized state, which we quantify by the area of corresponding closed orbits ($A$) in the ternary diagram (see inset). In comparison, if protection spillovers are not considered, there exists a critical value of $\theta=\theta_c=0.067$, which must be reached for oscillations to emerge \cite{szabo_jpa04}. Even if we use the same initial conditions in the classical version of the game, any initial deviations from the simplex will be damped and the system ultimately returns back to the center of the simplex for all $\theta<\theta_c$. Protection spillovers effectively lower this critical value less than 0.001. This effect can be understood if we consider the fact that vortices in the rock-paper-scissors game with protection spillovers are fixed in space and incessantly act as triggers of synchronization. If protection spillovers are absent, the vortices are no longer able to hold their position in a given spot of the population since there is nothing to prevent them from moving about. This movement acts as an additional source of noise that hinders the onset of synchronization, which can thus emerge only when the critical value of $\theta=\theta_c=0.067$ is exceeded.

\begin{figure}
\centerline{\epsfig{file=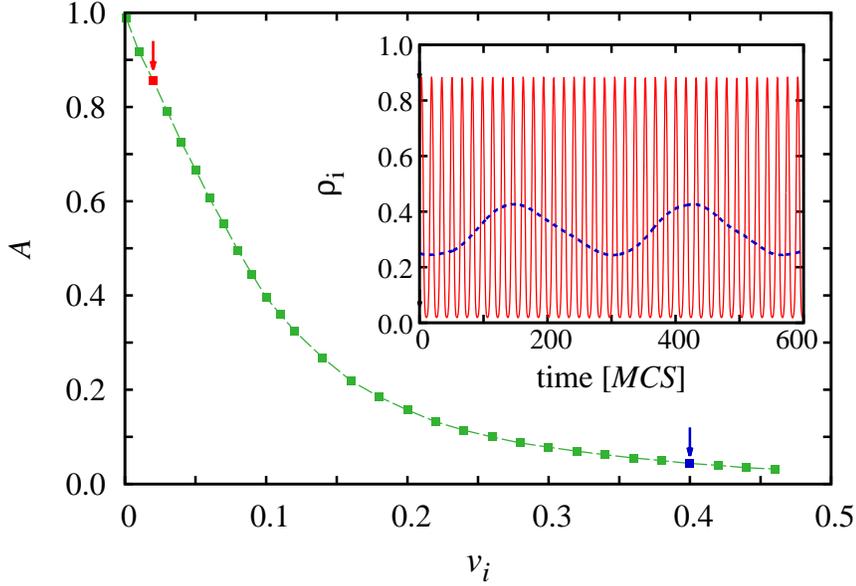,width=12cm}}
\caption{\label{shrink} The area of closed orbits ($A$) that correspond to stationary non-frozen states in the ternary diagram in dependence on the initial density of vortices ($v_i$), as obtained in the rock-paper-scissors game with protection spillovers in the regular random graph ($\theta=1$) limit with $L=10^6$ nodes. The inset shows the time evolution of the density of an arbitrary species, as obtained at a low initial density of vortices ($v_i=0.02$, solid red line) and at a high initial density of vortices ($v_i=0.4$, dashed blue line). The areas of these oscillations in the ternary diagram are marked by arrows with the same color in the main panel. These results demonstrate that as the initial density of vortices increases (as we approach the random initial state limit), not only does the amplitude (and hence the area in the ternary diagram; see main panel) decreases, but also that the frequency of oscillations decreases (see inset). Accordingly, the frozen state that is obtained at sufficiently high values of $v_i$ can be considered as the zero-amplitude infinite-period limit of the oscillations that are depicted in the inset.}
\end{figure}

Lastly, we also consider the $\theta \to 1$ limit, where the initial density of vortices in the population can be reasonably high without the system being destined to a frozen state (see Fig.~\ref{fixation_Q}). The obtained results are summarized in Fig.~\ref{shrink}, where the area of stationary orbits in the ternary diagram ($A$) is plotted in dependence on the initial density of vortices ($v_i$). Given that the value of $A$ decreases as we increase $v_i$, too many vortices evidently hinder the emergence of collective global synchronization. In fact, as the number of vortices that are fixed in space increases, so does the number of propagating waves. Sooner or later these waves meet and collide, which disturbs the emergence of synchronization. Moreover, the inset of Fig.~\ref{shrink} illustrates that not only does the amplitude of oscillations decreases, but also so does their frequency. Therefore, the frozen state in the $v_i \to 1$ limit can be considered as the zero-amplitude infinite-period limit of the depicted oscillations.

\section{Discussion}
Motivated by the preceding research of Kelsic et al. \cite{kelsic_n15}, who have shown that the opposing actions of antibiotic production and degradation enable stable coexistence, we have here studied the rock-paper-scissors game with protection spillovers in structured populations. Although the introduction of protection spillovers seems like a relatively minor amendment to the microscopic dynamics describing the rock-paper-scissors game, the consequences are quite spectacular. Depending on the initial conditions, it is certainly surprising how little of the original results that were obtained with the classical rock-paper-scissors game is recovered. While propagating waves dominate in the later case, we have shown that in the rock-paper-scissors game with protection spillovers the initial presence of vortices plays a key role. More precisely, we have shown that the spatiotemporal dynamics of the system is determined by the density of these vortices, which may ultimately transform to frozen states, to propagating waves, or to target waves with reversed propagation direction. Since the initial density of vortices might be controlled in experimental setups, our results thus reveal a feasible way of programming pattern formation in microbial populations.

We have also shown that annealed and quenched randomness in the interactions among species have a completely different impact on the dynamics of biodiversity. Importantly, this is not the case for the classical rock-paper-scissors game and related evolutionary games that are governed by cyclic dominance, where both sources of randomness have been shown to have the same impact in that they evoke synchronized oscillations among the competing species \cite{szabo_jpa04, szolnoki_pre04b}. For the rock-paper-scissors game with protection spillovers, our research reveals that, just like in the classical version of the game, quenched randomness facilitates collective synchronization in the population, which manifests as oscillations of strategy densities. Annealed randomness, however, favors the emergence of target waves, but with a reversed propagation direction where the vortices actually act as sinks for the wavefronts. To the best of our knowledge, we are unaware of other systems, either biological or chemical, that would exhibit this type of spatiotemporal dynamics, i.e., target waves with reversed propagation direction, effectively moving backward but looking like they are moving forward. In conclusion, we have shown that protection spillovers may fundamentally change the dynamics of cyclic dominance in structured populations, especially so under the impact of annealed randomness.

Our results have important and far-reaching implications. Protection spillovers are common in microbial populations, and it is in fact surprising that this has not received more attention in the past. In the standard setup, each bacterial species must protect itself from the toxin of its victim. For example, scissors protects itself from paper's toxin. A neglected aspect of this protection is that it may be non-excludable, meaning that protection may spill over to other species \cite{kelsic_n15, bergstrom_n15}. Such transitivity in protection may occur if a cell degrades the antimicrobials of a competing species by secreting enzymes that do the job externally, or by deactivating the competitor's antimicrobials once they have entered the cell \cite{yurtsev_msb13}. Regardless of the details, this reduces the concentration of the antimicrobial in the environment, thus giving rise to the here considered protection spillovers. As argued already by Kelsic et al. \cite{kelsic_n15}, these considerations have direct relevance for engineering multi-species microbial consortia and shed light on the dynamics of biodiversity in populations that are governed by cyclic dominance. Beyond microbial communities, cyclic dominance plays an important role also in marine benthic populations and plant systems, and the list of examples where the puzzle of biological diversity can be explained by cyclical interactions in the governing food webs is indeed impressively long and inspiring \cite{szolnoki_jrsif14, stouffer_s12}. Time will tell in which examples protection spillovers play a key role. Based on the presented results, however, it is certain that their impact is going to be a significant one, but also that reverse engineering this impact might be a difficult proposition.

\ack
This research was supported by the Hungarian National Research Fund (Grant K-101490), the Deanship of Scientific Research, King Abdulaziz University (Grant 76-130-35-HiCi), and by the Slovenian Research Agency (Grant P5-0027).

\section*{References}
\providecommand{\newblock}{}

\end{document}